\documentclass[fleqn,10pt]{wlscirep}
\usepackage[utf8]{inputenc}
\usepackage[T1]{fontenc}
\title{Tuning Magnetic Properties of Penta-Graphene Bilayers Through Doping with Boron and Oxygen}

\author[1]{Ramiro Marcelo dos Santos}
\author[1]{Wiliam Ferreira da Cunha}
\author[2]{Rafael Tim\'oteo de Sousa Junior}
\author[2]{William Ferreira Giozza}
\author[1,*]{Luiz Antonio Ribeiro Junior}
\affil[1]{Institute of Physics, University of Bras\'{i}lia , 70.919-970, Bras\'{i}lia, Brazil}
\affil[2]{Department of Electrical Engineering, University of Bras\'{i}lia 70919-970, Brazil}
\affil[*]{ribeirojr@unb.br}

\begin{abstract}
Penta-graphene (PG) is a carbon allotrope that has recently attracted the attention of the materials science community due to its interesting properties for renewable energy applications. Although unstable in its pure form, it has been shown that functionalization may stabilize its structure. A question that arises is whether its outstanding electronic properties could also be further improved using such a procedure. As PG bilayers present both sp$^2$ and sp$^3$ carbon planes, it consists of a flexible candidate for functionalization tuning of electromagnetic properties. In this work, we perform density functional theory simulations to investigate how the electronic and structural properties of PG bilayers can be tuned as a result of substitution doping. Specifically, we observed the emergence of different magnetic properties when boron was used as dopant species. On the other hand, in the case of doping with oxygen, the rupture of bonds in the sp$^2$ planes does not induce a magnetic moment in the material. 
\end{abstract}
\begin{document}

\flushbottom
\maketitle

\thispagestyle{empty}

\section*{Introduction}
The interest in two-dimensional extended nanostructures was further highlighted after graphene was obtained by the Novoselov group \cite{Novoselov666,geim}. The properties of graphene are truly unique \cite{Lin1294}, and a wide range of applications have been predicted for this carbon allotrope \cite{ming}. Although the overall success of graphene, there are still some limitations to be overcome regarding its optoelectronic applications, the main being its null bandgap \cite{Novoselov}. For that reason, other two-dimensional materials have been extensively tested. Among them, the transition metal dichalcogenides (TMDs) \cite{Qiu2013HoppingTT, PhysRevB.88.195313, 4420a96239b74951acceb77e0f47d4c5} and novel graphene-based allotropes \cite{zhang2015penta,wang2015phagraphene,wang2018popgraphene} stand out. However, graphene allotropes still hold the important advantage over TMDs, for instance, of being formed only by carbon atoms, which makes it easier to synthesize on a large scale. Therefore, it would be ideal to conceive a material that, while presenting semiconducting properties would still be based on graphene.

Having such a goal in mind, several theoretical and experimental studies have been conducted to achieve alternative solutions to monolayer graphene \cite{zhang2015penta,wang2015phagraphene,li2017psi,jiang2017twin,wang2018popgraphene,meng_PCCP,toh_2020,joo_SciAdv,falcao_JCTB}. For instance, studies concerning graphene bilayer have concluded that its structural properties can be altered from a suitable layer stacking engineering \cite{PhysRevLett.99.256802,PhysRevB.98.235402,Bistritzer12233}. In one of these studies, it was obtained that the rotation of one graphene layer over another leads to the displacement of the Dirac cones, but no gap opening was observed \cite{PhysRevLett.99.256802}. The conclusion is that layer stacking still need to be associated with other procedure if a graphene-based system with semiconducting properties is to arise. So far, the main option relied on substitutional doping \cite{PhysRevB.81.235401, PhysRevLett.101.086402, Gweon, Ohta951} on graphene structure. Density functional theory calculations were performed to study the doping of boron (B) and nitrogen (N) in graphene bilayers \cite{fujimoto_SS}. The results revealed that for doping with B (N), a gap opening from 0.11 to 0.32 eV (0.09 and 0.30 eV) was obtained, being these values depend on the type of bilayers packing \cite{fujimoto_SS}.

Another approach is to study other possible arrangements: different, and yet based on graphene. Among them are carbon nanotubes  \cite{PhysRevB.96.075133}, nanoribbons  \cite{doi.org/10.1038/nature07919}, nanoscrolls  \cite{doi:10.1021/nl900677y}, and novel graphene-based allotropes, such as popgraphene \cite{wang2018popgraphene}, phagraphene \cite{wang2015phagraphene}, and penta-graphene \cite{zhang2015penta} (PG). In its pure form, the PG structure is meta-stable but presents an almost direct bandgap $\sim 2.4$ eV, which is interesting for optoelectronic applications. Its structure is similar to the pentagonal tiles of Cairo, where the carbon atoms occupy three types of planes: a plane of sp$^3$ carbons sandwiched by two planes of sp$^2$ carbons \cite{zhang2015penta}. Such a structure allows more degrees of freedom for doping when contrasted to graphene. As a result, different ways of modulating its electronic properties can be obtained \cite{dos2020tuning}. Particularly, it can be reasoned that the flexibility of such material makes it a promising candidate to fulfill the role of the carbon allotrope with semiconducting-like properties. This fact is especially true if one considers the two promising factors on PG: considering bilayers and substitutional doping.

It is well known that the PG bandgap increases with an increase in oxygen doping concentration \cite{li_RSCADV,zhang_CMS,mi_PHYSOPEN}. Such a trend is attributed to the up-shift of the conduction band minimum during the oxidation process \cite{li_RSCADV}. Since oxidized PG exhibits a large bandgap, it can be considered as a good alternative for the conception of new dielectric layers in electronic devices \cite{zhang_CMS}. When it comes to doped carbon-based nanostructures, boron is the most often used dopant once its similar electronic structure and size to carbon allow its incorporation into the carbon-based substrate with minimal strain and changes in the lattice arrangement \cite{zhang_NANOMATMDPI,berdiyorov_JPCM,krishnan_PCCP,sathishkumar2019boron,chigo2016long,anota2014studies}. Moreover, the boron‐doped PG can effectively decompose H2 molecules into two H atoms, which is an interesting feature for energy conversion and storage applications \cite{sathishkumar2019boron}. In this sense, oxygen- and boron-doped PG systems must be further studied to propose new routes in developing PG-based devices. 

Herein, motivated by the fact that PG has more degrees of freedom to modulate its electronic properties than graphene, we have explored its electronic properties in doped bilayers by carrying out state of art electronic structure calculations on the Density Functional Theory (DFT) level of theory. We studied the effects of boron and oxygen as substitutional dopants in the different hybridization planes of the PG structure. It was observed that the introduction of oxygen atoms in the sp$^2$ plane leads to the breaking of bonds and subsequently no magnetic properties take place. Remarkably, boron doping in the sp$^3$ plane, in turn, gives rise to a significant magnetization which was not observed for the cases in which this dopant was introduced in the sp$^2$ plane. As mentioned above, PG packing allows us to have extra degrees of freedom for doping. This feature is interesting since in PG monolayers there are two sp$^2$ planes that can interact with the external environment. This, in turn, creates another degree of freedom for doping and, consequently, for modulation of its electronic properties. 

\section*{Results}
\begin{figure}[!htb]
\centering
\includegraphics[scale=0.38]{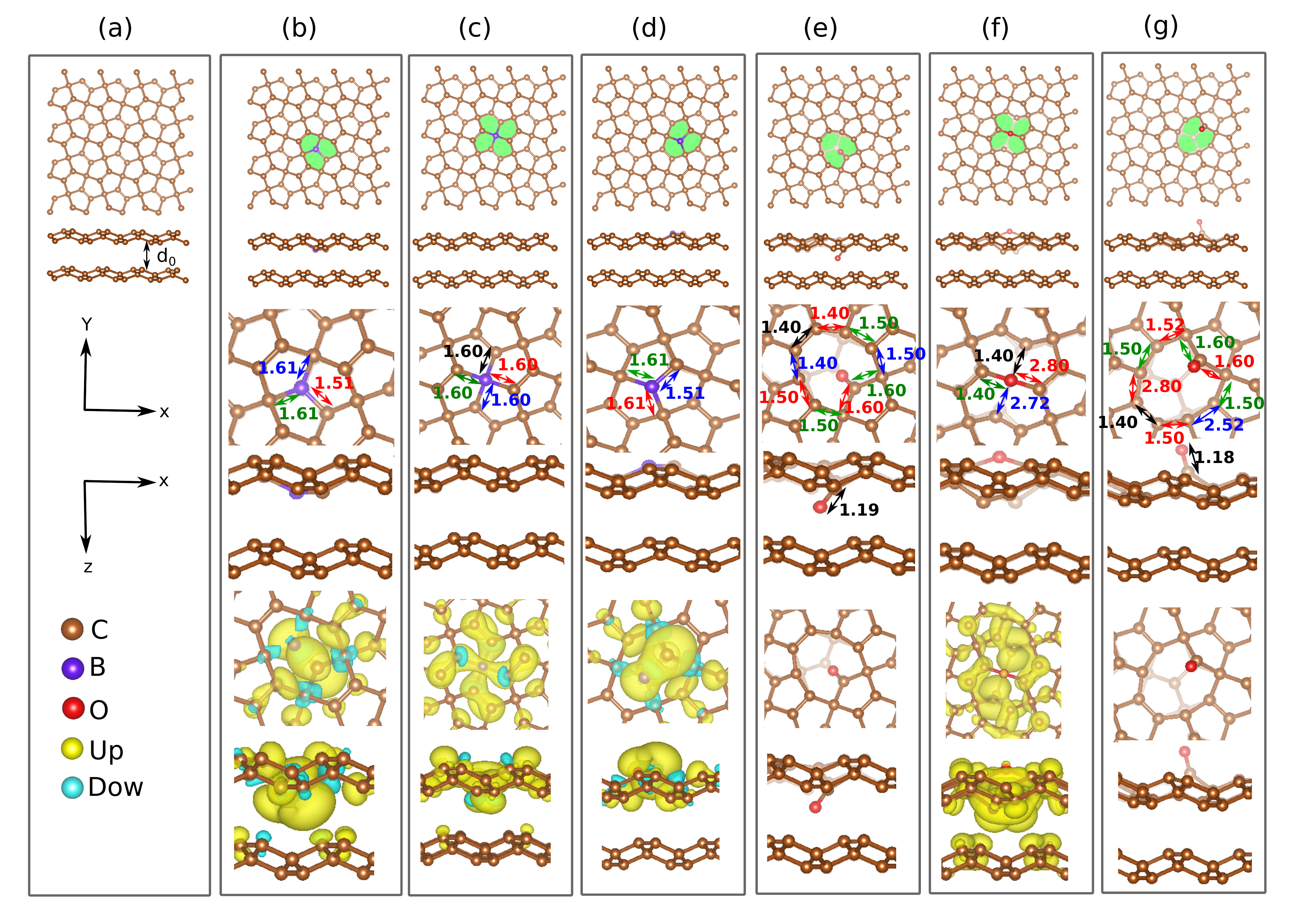}
\caption{Schematic representation of optimized PG bilayers doped with boron and oxygen. The middle panels show the bond lengths nearby the dopant site. The bottom panels illustrate the charge density distribution, $\rho(Up)-\rho(Down)$, obtained by adopting an isovalue of 0.001.}
\label{fig1}
\end{figure}

We begin our discussions by presenting the structural properties and the charge localization profiles for the model doped PG bilayers studied here. In this sense, Figure \ref{fig1} shows the schematic representation of optimized PG bilayers doped with boron and oxygen (top panels), the bond lengths on the neighboring of the doping sites (middle panels), and charge density distribution of the system (bottom panels). Figure \ref{fig1}(a) presents the optimized geometry of the two layers without defects, i.e. before the doping procedure took place. Figures \ref{fig1}(b-d) represent the substitutional doping with boron, whereas Figures \ref{fig1}(e-g) account for the oxygen-based functionalization. In the doping mechanism adopted here, the PG plane at the bottom is always pristine. The second PG layer on top receives the dopant in three distinct channels: (dopant-sp$^2$-out) when the dopant is placed in the $sp^2$ plane above the interlayer region interacting with the vacuum, (dopant-sp$^2$-in) when the dopant is placed in the $sp^2$ plane within the interlayer region interacting with the PG plane at the bottom, and (dopant-sp$^3$) for the doping in the sp$^3$ plane. For these nomenclatures, ``dopant'' stands for oxygen (O) or boron (B) atom. Importantly, some theoretical works have studied the structural stability of oxygen- and boron-doped PG layers \cite{sathishkumar2019boron,zhang2017remarkable}. In a reactive molecular dynamics study, the results have revealed that oxygen-doped PG layers present remarkable enhancement in failure stress and strain when contrasted with pristine PG layers \cite{zhang2017remarkable}. Moreover, it was recently demonstrated, by using DFT calculations, that boron-doped PG layers are structurally stable when concentrated with pristine PG layers \cite{sathishkumar2019boron}.    

Figures \ref{fig1}(b), \ref{fig1}(c), and \ref{fig1}(d) illustrate the cases B-sp$^2$-in, B-sp$^3$, and B-sp$^2$-out, respectively. One can observe that the bond lengths between $B-C$ slightly deviate from that of the bond length $C-C$ in the pristine monolayer, which is 1.57 \AA. These bond lengths ($B-C$) assume minimum and maximum values of 1.51 to 1.61 \AA for all the cases of doping with B. As for the oxygen doping picture, represented by Figures \ref{fig1}(e-g), we observe a greater deformation around the doping site. Figures \ref{fig1}(e) and \ref{fig1}(g) show the result of doping the sp$^2$ planes. It is possible to note a tendency of carbonyl formation with the elevation of the oxygen atom by the distance of 1.18 and 1.19 \AA from the carbon atom in the respective sp$^3$ plan. With the elevation of the oxygen from the sp$^2$ plane, the formation of a vacancy is observed with bond lengths in the edges whose sizes vary from 1.40 to 1.60 \AA for the case O-sp2-in and O-sp2-out. In the case O-sp3 (Figure \ref{fig1}(f)), a tendency of the oxygen to leave the plane is also observed, as the bond length of 1.40 to 1.72 \AA is achieved between the first neighbors of the dopant. 

The bottom panels of Figure \ref{fig1} provide a representation of the charge density wrapped around the dopant. The yellowish cloud stands for the density due to up spin electrons, whereas the blueish on of the down spin electrons. The net charge observed is responsible for the magnetic moment, which characterizes a magnetization due to the presence of the dopant. The values of the magnetic moment for each structure are listed in Table \ref{tab1}, to be discussed later. It is observed that this charge concentration is more effective in the case of doping with boron. On the other hand, no spontaneous magnetization was observed in the O-sp2-in and O-sp2-out cases. This behavior suggests that magnetization in the case of doping with oxygen is not a direct consequence of doping, but rather of the deformation of the geometry that the dopant produces. The sp$^3$ plane has two dangling carbons with a distance of 2.80 and 2.72 \AA from the oxygen atom. This distance makes the $\pi$-electrons of these two atoms to contribute to the effective magnetization of the bilayer, whose magnetic moment is 2.00 Bohr Magneton. In the O-sp2-in and O-sp2-out cases, the non-concentration of charge in the oxygen atom, which makes two connections in the plane, suggests a double bonding with the carbon, which characterizes a carboxyl. The resulting bond configuration for the carbon atoms in the edge keeps the symmetry of the vacancy.

\begin{table}[!htb]
\centering
   \begin{tabular}{|l|l|l|l|l|l|l|}
\hline
& B-sp2-in & B-sp3 & B-sp2-out& O-sp2-in& O-sp3 & O-sp2-out \\
\hline
  E$_{form}$ (eV) & -9.54 & -9.55 & -9.39 & 9.60 & -9.59 & -9.41 \\
  E$_{coh}$ (eV) & -8.38 & -8.37 & -8.37 & -8.42 & -8.48 & -8.42 \\
  \textbf{m} ($\mu_B$)& 1.00 & 0.93 & 1.00 & 0.00 & 2.00 & 0.00 \\  
   d$_0$ (\AA) & 3.12& 2.50 & 2.60 & 2.70 & 2.60 & 2.60 \\
\hline
\end{tabular}
\caption{Formation energy, cohesion energy, magnetic moment and distance between the two layers.}
\label{tab1}
\end{table}

The first row of Table \ref{tab1} presents the formation energy, which is defined as: $E_{form} = E_b-(E_{mp}+E_{md})$, where $E_b$ is the total energy for each doped bilayer, $E_{mp}$ the total energy of the monolayer without defect, and E$_{md}$ total energy of the doped bilayer. One can observe that the energy cost for the formation of the bilayer is approximately the same for the two types of doping, i.e., -9.39 to -9.60 eV for boron and oxygen, respectively. In the second row of this table, we present the cohesion energy that is given by $ E_{coh} = \frac{E_{total}-(E_{C}N_{C}+E_{X}N_{x})}{N_{total}}$, where $E_{C}$ is the total energy of a carbon atom and $N_{C}$ is the number of carbon atoms of the bilayer. $E_{x}$ and $N_{x}$ are the total energies and number of doping atoms respectively, where $X$ stands for B or O. $N_{total}$ is the total number of atoms in the bilayers. The cohesion energy has values similar for all systems, suggesting the same level of cohesion for the two types of dopants. In the third row, we present the magnetic moment of each bilayer. One can realize that for the O-sp3 case, the value of the magnetic moment is zero, which is in agreement with what was already discussed of the net charge density. In the fourth row, the distances $d_0$ between the two layers are shown. After optimizing the geometry of the bilayers, these distances remained larger for the two both B-sp2-in and O-sp2-in doping cases, compared to the others that were between 2.50 and 2.60 \AA, while for B-sp2-in and O-sp2-in is 3.12 and 2.70 \AA, respectively. These results suggest that the dopant has a significant contribution to the interactions between the two layers.

\begin{figure}[!htb]
\centering
\includegraphics[scale=0.78]{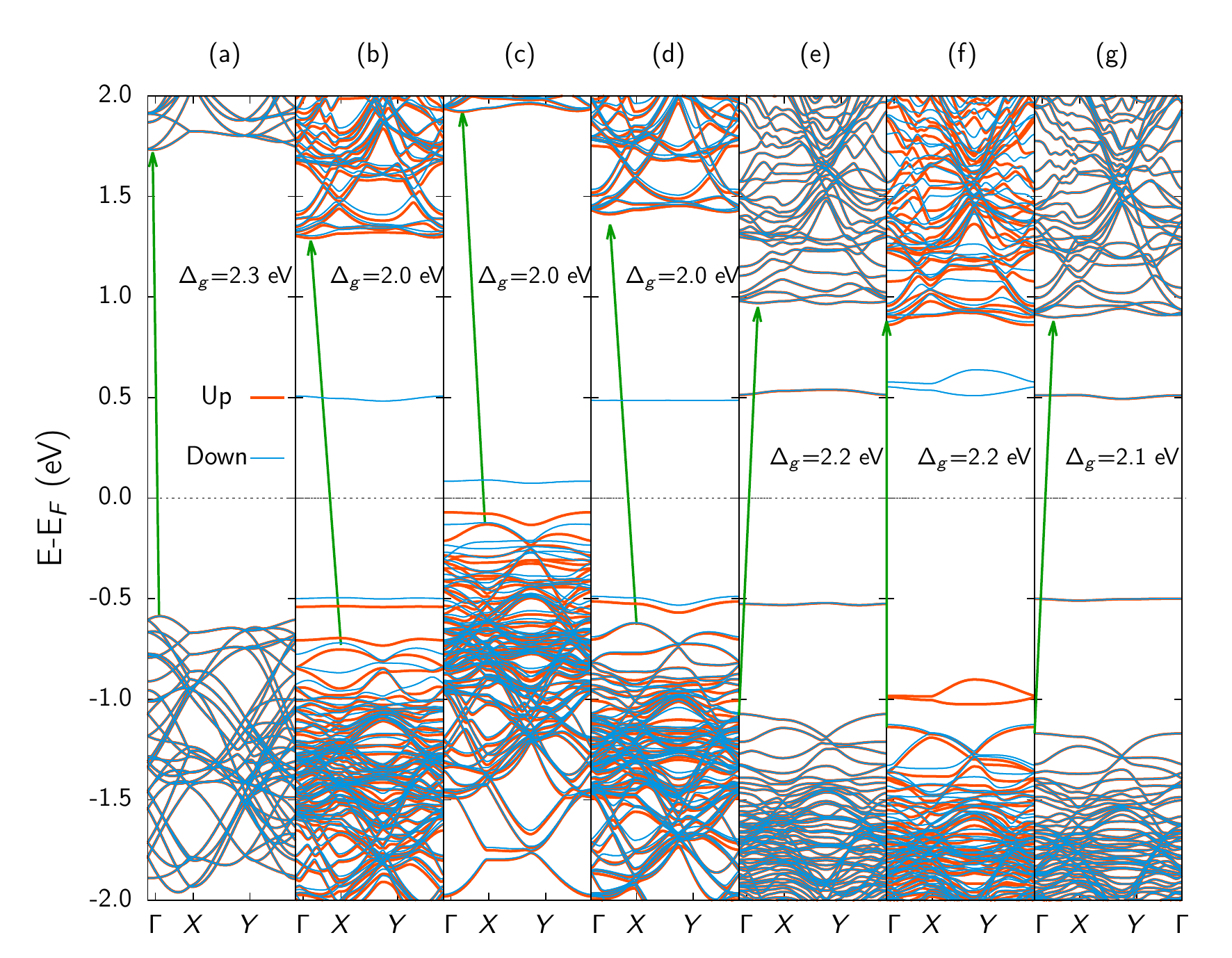}
\caption{Band structure for the doped PG bilayers studied here. (a) Pristine srtucture (b) B-sp2-in (c) B-sp3 (d) B-sp2-out. In (e), (f), and (g) O-sp2-in, O-sp3, and O-sp2-out are respectively illustrated.}
\label{fig2}
\end{figure}

Figure 2 presents the band structure of the doped bilayer bands. For the pure bilayer (Figure \ref{fig2}(a)), one can note that an almost indirect gap of 2.3 eV takes place, a value approximately equal to that found for the PG monolayer \cite{zhang2015penta}. In the case of doping with boron, we observe a reduction of the bandgap to 2.0 eV, which remained indirect. For the two cases of sp$^2$ doping (Figures \ref{fig2}(b) and \ref{fig2}(d)), it occurs the emergence of states in the middle of the bandgap: a downstate above the Fermi level, one up and other down states, below the Fermi level. These states characterize the magnetization of the system. For the sp$^3$ (Figure \ref{fig2}(c)), the two states in the middle of the bandgap are symmetrical, with spin up and down above and below the Fermi level, respectively. In the case of oxygen doping, we also observed the appearance of the states in the middle of the Fermi level: these states are symmetrical concerning the Fermi level and have no spin degeneration for the O-sp2-in and O-sp2-out cases (Figures \ref{fig2}(e) and \ref{fig2}(g)). In these configurations, the bandgap was reduced to 2.2 and 2.1 eV, respectively. For the O-sp3 case (Figure \ref{fig2}(f)), the bandgap is also 2.2 eV and with the two symmetrical states about the Fermi level, with spin up below and down above the Fermi Level.

\begin{figure}[!htb]
\centering
\includegraphics[scale=0.78]{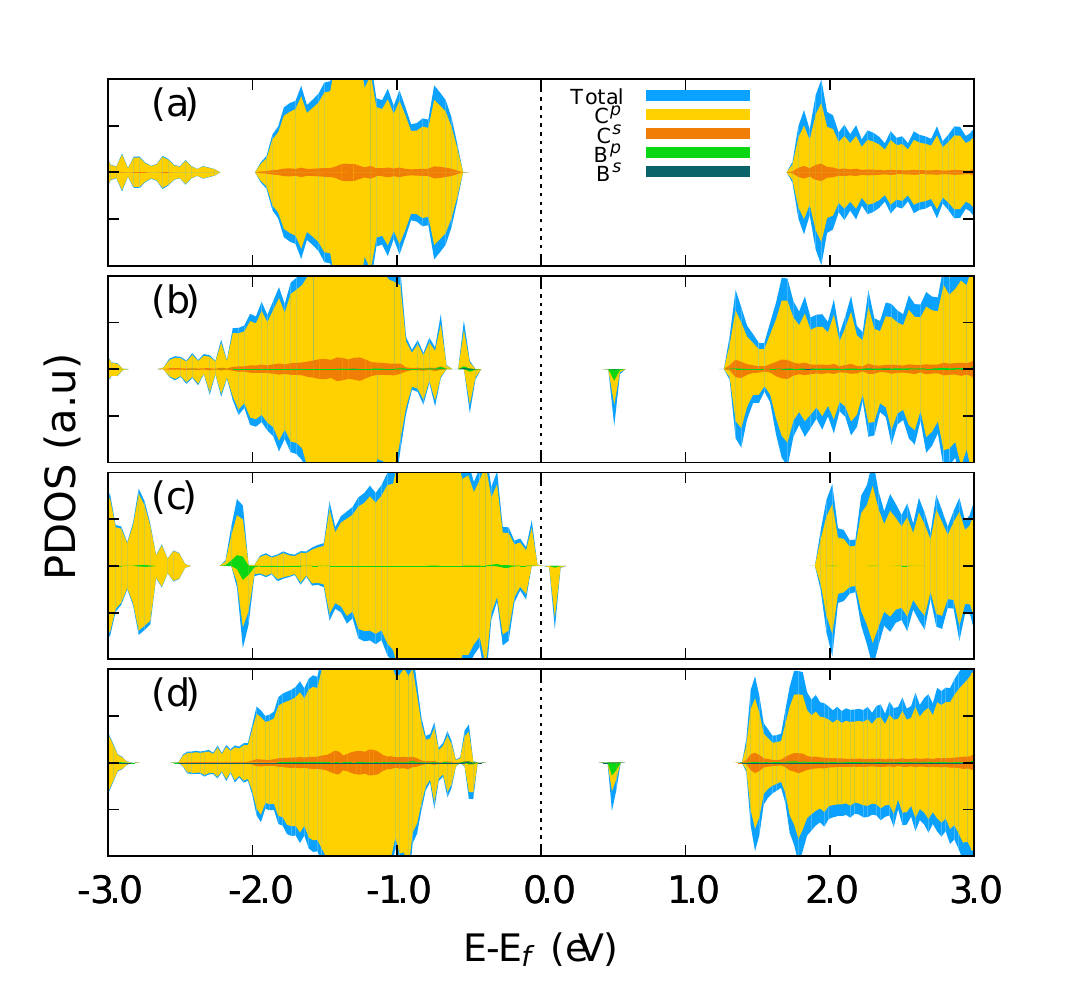}
\includegraphics[scale=0.78]{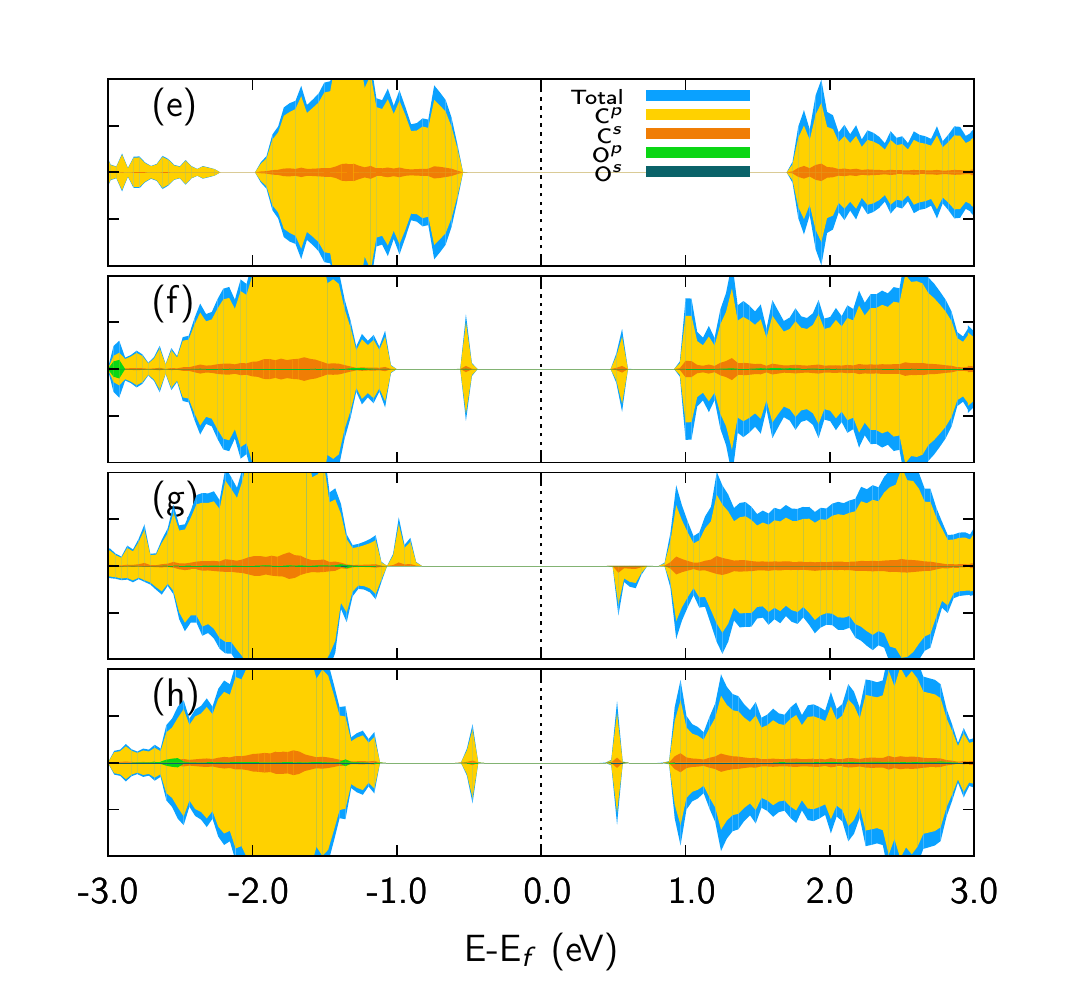}
\caption{Projected density of states (PDOS) for the doped PG bilayers studied here. (a) Pristine srtucture (b) B-sp2-in (c) B-sp3 (d) B-sp2-out. In (e), (f), and (g) O-sp2-in, O-sp3, and O-sp2-out are respectively illustrated. }
\label{fig3}
\end{figure}

Finally, Figure \ref{fig3} presents the projected density of states (PDOS). In the cases of boron doping, one can realize that for the pristine bilayer (Figure \ref{fig3}(a)) the most significant contribution to the formation of the bands is for the $p$ states of $sp^2$ carbon atoms. The Fermi level is closer to the valence bands, which characterizes a $n$-type semiconductor. As for the B-sp3 cases (Figures \ref{fig3}(b) and \ref{fig3}(d)), a slight contribution of $p$ orbitals of boron at the first peak in the middle of the bandgap (above the Fermi level) is observed. Still, for these two cases, we observed that the Fermi level is closer to the valence band, which characterizes a $n$-type semiconductor. Regarding the B-sp3 case (Figure \ref{fig3}(c)), there is no significant contribution of boron to the states near the Fermi level and the structure behaves as a $n$-type semiconductor, once the Fermi level is touching the top of the valence band. In the case of oxygen doping (Figures \ref{fig3}(e-h)), we did not observe any significant contribution from O atoms. In the O-sp2-in case (Figure \ref{fig3}(f)), the Fermi level is in the middle of the bandgap with peaks symmetrically localized regarding it for both spin channels, which characterizes the non-magnetization of this material. In the O-sp3 case (Figure \ref{fig3}(g)), we have observed that the Fermi level is slightly displaced towards the conduction states, which characterizes a $p$-type semiconductor. It is also observed antisymmetric peaks in relation to the spin, which stands for a significant magnetic moment of this structure. For the O-sp2-out case (Figure \ref{fig3}(h)), a slight displacement of the Fermi level near the conduction band was also observed, characterizing a $p$-type semiconductor. It is observed that the peaks closest to the Fermi level are symmetrical in relation to the spin states, proving the non-magnetic behavior of the material.

\section*{Methods}
To investigate the electronic structure of doped PG bilayers, we used DFT calculations as implemented in the SIESTA software \cite{PhysRevB.53.R10441,Portal}. It makes use of a numerical base to expand the wave functions of the many atoms system. In the present work, it was used the DZP basis set \cite{PhysRev.136.B864, PhysRev.140.A1133}. As for the functional approximation, it was considered the generalized approximation of the gradient proposed by Perdew, Burke and Ernzerhof (GGA/PBE) + vDW  \cite{PhysRevLett.77.3865}, which is built from the expansion of the second-order density gradient. Pseudopotentials parameterized within the Troullier-Martins formalism are also considered \cite{PhysRevLett.48.1425}. This approximation is of fundamental importance for the description of the magnetic and electronic properties of materials composed of atoms with many electrons. All calculations were performed considering spin polarization. To calculate the bands and state densities, an MPK mesh of 15 x 15 x 1  is used \cite{PhysRevB.13.5188}. A mesh cut of 200 Ry is chosen as a parameter for our calculations \cite{PhysRevB.66.205101}. The forces converged until reaching a minimum value of 0.001 eV/\AA. In order to ensure a good compromise between the accuracy of our results and the computational feasibility, the tolerance in the density of the matrix and in the total energy was set at 0.0001 and 0.00001 eV, respectively. Importantly, this set of parameters were used recently to study other carbon-based lattices \cite{dos2020tuning,dos2019defective,dos2019electronic,da2019tuning}. 

\section*{Conclusion}
In summary, we carried out DFT calculations to investigate the influence of boron and oxygen doping on the electronic properties of PG bilayers. Our findings showed that the difference between dopant on the sp$^2$ and sp$^3$ planes have a significant impact on the magnetic properties of oxygen doping. As for the case of doping with boron, it was observed a spontaneous magnetization in the system. This is due to the fact that boron contains only one electron in the orbital valence $2p$, whereas the substituted carbon has a pair of electrons in this same orbital. Therefore, the $C-B$ bond results in a magnetic polarization due to electronic covalence, leading to the emergence of two electrons with the same spin and one electron with the opposite spin. This effect is characterized by three flat energy levels that appear in the middle of the bandgap for the case of boron doping in the sp$^2$ planes. Such an effect is not observed in the case of doping with oxygen due to bond breaking. Regarding the effect on the pristine layer, there was no significant difference in its electronic properties. In the cases of boron doping, the resulting system characterizes a $n$-type semiconductor. On the other hand, for the oxygen doping in the sp$^3$ plane, we have observed that the Fermi level is slightly displaced towards the conduction states, which characterizes a $p$0-type semiconductor. Since the electronic structure of PG bilayers present extra doping channels and can be easily tuned by doping its structure with just a single atom, they can represent an interesting alternative for replacing graphene in some optoelectronic applications.   

\bibliography{references.bib}

\section*{Acknowledgements (not compulsory)}

The authors gratefully acknowledge the financial support from Brazilian Research Councils CNPq, CAPES, and FAPDF and CENAPAD-SP for providing the computational facilities. L.A.R.J. and W.F.G. gratefully acknowledge, respectively, the financial support from FAP-DF grants 0193.001.511/2017 and 00193.0000248/2019-32. L.A.R.J. gratefully acknowledges the financial support from CNPq grant 302236/2018-0. R.T.S.J. gratefully acknowledge, respectively, the financial support from CNPq grant 465741/2014-2, CAPES grantsWiliam and 88887.144009/2017-00, and FAP-DF grants 0193.001366/2016 and 0193.001365/2016.

\section*{Author contributions statement}

R.M.S and W.F.C. ran the calculations. R.M.S. and W.F.G. built the graphics. L.A.R.J., W.F.C., W.F.G., and R.T.S.J. interpreted the results and wrote the paper. All the authors were responsible for discussing the results.

\section*{Competing interests}
The authors declare no competing interests.

\end{document}